  \documentstyle[11pt,psfig]{article}
  
\newcommand{\be}{\begin{equation}}
\newcommand{\ee}{\end{equation}}
\newcommand{\bea}{\begin{eqnarray}}
\newcommand{\eea}{\end{eqnarray}}
\newcommand{\beann}{\begin{eqnarray*}}
\newcommand{\eeann}{\end{eqnarray*}}
\newcommand{\beasn}{\begin{sneqnarray}}
\newcommand{\eeasn}{\end{sneqnarray}}
\newcommand{\bref}[1]{(\ref{#1})}
\newcommand{\eps}{\epsilon}

\newcommand{\der}[2]{\frac{\partial #1}{\partial #2}}

\newcommand{\NPB}[3]{{\sl Nucl. Phys.} {\bf B#1} (19#2),  {#3}}
\newcommand{\PRD}[3]{{\sl Phys. Rev.} {\bf D#1} (19#2),   {#3}}
\newcommand{\PLB}[3]{{\sl Phys. Lett.} {\bf #1B} (19#2),  {#3}}
\newcommand{\PRL}[3]{{\sl Phys. Rev. Lett.} {\bf #1} (19#2), {#3}}
\newcommand{\ZFP}[3]{{\sl Zeitsch. f. Physik} {\bf C#1} (19#2),  {#3}}

\newcommand{\PRep}[3]{{\sl Phys. Rep.} {\bf #1} (19#2),  {#3}}
\title{{\bf M.C.R.G. Study of Fixed-connectivity Surfaces}}

\author{{{\sc D. Espriu}\thanks{espriu@greta.ecm.ub.es}
	   \ and {\sc A. Travesset}\thanks{alex@greta.ecm.ub.es}}\\
       \llap{}%
       \small{ Departament d'Estructura i Constituents
               de la Mat\`eria}\\
       \small{ Universitat de Barcelona} \\
       \small{ and} \\
       \small{ Institut de F\'\i sica d'Altes Energies}\\
       \small{ Diagonal, 647}\\
       \small{ E-08028 Barcelona} }

\date{}

\begin{document}

\maketitle

\begin{abstract}
We apply Monte Carlo Renormalization group to the crumpling transition
in random surface models of fixed connectivity. This transition
is notoriously difficult to treat numerically.
We employ here a Fourier accelerated Langevin
algorithm in conjunction with a novel blocking procedure in momentum
space which has proven extremely successful in $\lambda\phi^4$.
We perform two successive renormalizations
in lattices with up to $64^2$ sites.
We obtain a result for the
critical exponent $\nu$ in general agreement with
previous estimates and similar error bars, but with much less
computational effort. We also measure with great accuracy
$\eta$. As a by-product we
are able to determine the fractal dimension $d_H$ of random surfaces
at the crumpling transition.
\end{abstract}

\vfill
\vbox{
 UB-ECM-PF-95/20\hfil\null\par
 December 1995\hfil\null\par}

\clearpage

%%%%%%%%%%%%%%%%%%%%%%%%%%%%%%%%%%%%%%%%%%%%%%%%%%%%%%%%%%%%%%%%%%%%%%%%

\section{Introduction}

The idea of implementing Wilson renormalization group \cite{WK} in
numerical simulations was advocated by Swendsen in \cite{SW1},
where it was successfully applied to Ising spins.
In this paper we shall use this approach to study the
crumpling transition for
fixed-triangulation surfaces.

In \cite{polyakov} it was proposed to add to
the usual area or
Nambu-Goto term \cite{nambu} a piece proportional
to the extrinsic curvature.
The model has then a nontrivial $\beta$ function, the coupling being
asymptotically free. An auxiliary metric can be introduced and
the model interpreted as corresponding to $2d$
gravity coupled to {\em interacting} scalar fields.
If one  does not integrate over the auxiliary  $2d$ metric the model
admits a direct lattice transcription in terms of surfaces
with fixed connectivity; i.e. crystalline surfaces\cite{oldsim},
\cite{Kan}.
It is remarkable that for some intermediate
value of the coupling $\kappa$ the model shows a
second order phase transition.
This transition will be the object of our interest.

Whenever one simulates a statistical system near a
critical point one immediately faces the problem of
critical slowing down.
Critical slowing
down is a consequence of the fact that near criticality the low momentum
evolve very slowly with local algorithms.
An algorithm that partially beats
slowing down was introduced in \cite{WL1} and uses a
Fourier accelerated Langevin algorithm (FALA).
This algorithm permits different
updatings for different modes hence
partially beating critical slowing down. FALA also facilitates
a direct implementation of MCRG in momentum space
by performing Kadanoff transformations that consist in decimations
of the higher momenta, close to the original Wilson spirit\cite{WK}.
A very successful
application of these ideas to $\lambda\phi^4_3$ was presented
in \cite{ET1}.

The organization of this paper is as follows. In section 2 we will
give a general overview of the formal points of the method. In section
3 we will make a detailed discussion about the algorithm and
the problems that need be circumvented in its applications.
In section 4 we survey the previous work concerning the
crumpling transition and fix some conventions.
In section 5 we analyze extensively the results from
our simulation and give estimates for the critical exponents. In
section 6 we will point out some conclusions and possible extensions of
this work.

\section{The Method}

The method we will use has been applied
successfully to scalar theories\cite{ET1}. Preliminary results dealing
with the problem of surfaces have already appeared in\cite{ET2}. In this
section we will give a quick overview.
Our starting point is a bare action
\be
S=\sum_{\alpha} \lambda^{b}_{\alpha} O_{\alpha}^{b}
\ee
After applying a renormalization group
transformation with a scale $t$ we
end up with a similar action, but with renormalized couplings
\cite{WK}:
$\lambda_{\alpha}^{r}=F\left(\lambda_{\alpha}^{b} \right)$.
We can linearize $F$ in the vicinity of the fixed point
$$
\lambda_{\alpha}^{r}-\lambda_{\alpha}^{*}=
   \sum_{\beta} T_{\alpha \beta} \left(
   \lambda_{\beta}^{b}-\lambda_{\beta}^{*} \right) ,
$$
\be
T_{\alpha \beta}={
\left( \der{\lambda_{\alpha}^{r}}{\lambda_{\beta}^{b}} \right)}_
{\lambda^{b}=\lambda^{*}}.
\ee
The largest eigenvalue $\lambda_{h}$ of $T_{\alpha \beta}$ will give
the critical exponent $\nu$ according to the formula
$\nu=\log t/\log \lambda_{h}$. Using
the chain rule we get
\be
\der{\langle O_{\lambda}^{r} \rangle}{\lambda_{\beta}^{b}}=
\sum_{\alpha}{ \left(\der{\lambda_{\alpha}^{r}}
{\lambda_{\beta}^{b}} \right)
      \der{\langle O_{\lambda}^{r} \rangle}{\lambda_{\alpha}^{r}}},
\ee
with
\be
\der{\langle O_{\gamma}^{r} \rangle}{\lambda_{\beta}^{b}}=
{\langle O_{\gamma}^{r} O_{\beta}^{b} \rangle}-
{\langle O_{\gamma}^{r} \rangle \langle O_{\beta}^{b} \rangle},
\qquad
\der{\langle O_{\gamma}^{r} \rangle}{\lambda_{\beta}^{r}}=
\langle O_{\gamma}^{r} O_{\beta}^{r} \rangle-
\langle O_{\gamma}^{r} \rangle \langle O_{\beta}^{r} \rangle.
\label{correlators}
\ee
These correlators
and, in turn, $T_{\alpha \beta}$ and $\nu$ can be numerically
computed.

A renormalization group transformation consists of a Kadanoff
transformation that eliminates short distance degrees of freedom,
and a rescaling of the fields.
The Kadanoff transformation
is largely arbitrary. The one we will use here will be a decimation in
momentum space. At each step we will simply discard the high $p$ modes,
half of them in each direction.
This transformation can be implemented with a null cost in computer
time and it has many advantages, as it will be evident in what follows.
The rescaling of the fields is required in order that the
normalization of a given operator (for example, the kinetic
term in a scalar field theory) is
preserved after each transformation. That
amounts to redefining\cite{WK,ET1}
\be
 \phi^{r}=\zeta \phi^b \quad \quad \quad  \zeta=1/2^{(d+2-\eta)/2}
\ee
The quantity $\eta$ is, if computed at the fixed point, the conformal
dimension of the field. In the case of surfaces is directly
related to the Hausdorff dimension (see section 4).
$\eta$  will be obtained selfconsistently by demanding that
bare and renormalized observables should
coincide at the fixed point.
This delivers very precise values for $\eta$.
Once $\nu$ and $\eta$ are determined we can get the rest of them
through
hyperscaling relations (see section 4).

It remains to select an efficient algorithm to compute the
correlators \bref{correlators}.
The crumpling transition has extremely long
autocorrelation times near the critical point.
autocorrelation times are surprisingly long (up to $10^6$ local
updates in a $128\times 128$ system!) and that the asymptotic
behaviour for the specific heat does not set in until the size of
the system is relatively large.
There is a whole industry of
algorithms to overcome the problem of critical slowing down\cite{sok1}.
For the problem at
hand, the crumpling transition,
FALA\cite{WL1} has been successfully employed\cite{Wh1,Wh2,Wh3,ET2}.
Its advantages are twofold: we
partially beat critical slowing down and we can
implement our Kadanoff transformation easily.

The implementation of the algorithm is not completely straightforward,
however. It is thus
worth to pause for one second and briefly digress on its
foundations.
We begin by constructing a random
walk in configuration space via the stochastic Langevin equation
\be
\phi(x,t_{n+1})=\phi(x,t_{n})+\Delta\phi(x,t_n),
\label{lang1}
\ee
\be
\Delta\phi(x,t_n)=-\Delta t \sum_{y}\eps(x,y) \der{S}{\phi(y,t_n)}+
\sqrt{\Delta t}\ \eta(x,t_n),
\label{lang2}
\ee
where $\eta(x,t)$ is a gaussian noise which satisfies
\be
{\langle \eta(x,t) \eta(y,t') \rangle}_{\eta}=2\delta_{t,t'}\eps(x,y).
\ee
The choice
$\eps(x,y)=\delta_{x,y}$
corresponds to a local updating which is more or less equivalent
to the familiar Metropolis algorithm, with a dynamical
exponent $z=2$. It has been used
in random surfaces in \cite{am1}, but suffers from long
autocorrelation times and the same difficulties that will
be discussed in the next section.

If we instead make the choice
\be \eps(x,y)=\sum_{p} \exp(-ip(x-y)) \eps(p)
\label{eps1}
\ee
with a wise election of $\epsilon(p)$ we can accelerate separately
the different Fourier modes.
The equations after a simple
rescaling read
\be
\phi(p,t_{n+1})=\phi(p,t_n)-\Delta t \eps(p){\cal F}\der{S}{\phi(y,t_n)}
+\sqrt{\Delta t \eps(p)}\eta(p,t_n),
\ee
\be
\langle \eta(p_1,t) \eta(p_2,t') \rangle=2\delta_{t,t'}\delta_{p1,-p2},
\ee
where ${\cal F}$ denotes the Fourier transform.
In the previous equations the modes are
manifest; by choosing a function
$\eps(p)$ which enhances the low $p$-modes
critical slowing down can be reduced as the longer equilibration times
that low $p$-modes need are roughly made up by correspondingly
larger updates.

From the previous Langevin equation one can
derive the Fokker-Planck equation
\be
P(\phi,t_{n+1})-P(\phi,t_{n})=\sum_{n=1}^{n=\infty}\sum_{x_1,..,x_n}
\der{}{\phi(x_1)} \der{}{\phi(x_2)}..\der{}{\phi(x_n)} \left(
\Delta_{x_1,x_2,..,x_n} P(\phi,t_n) \right),\label{FKP1}
\ee
\be
\Delta_{x_1,x_2,..,x_n}=\frac{(-1)^n}{n!} \langle \Delta\phi(x_1,t_n)
\Delta\phi(x_2,t_n)...\Delta\phi(x_n,t_n) \rangle,
\label{FKP2}
\ee
whose stationary solution is, in powers of $\Delta t$,
\be
P(\phi)=\exp\left(-S(\phi)+\frac{\Delta t}{2}\sum_{x,y}\eps(x,y)\left(
\frac{\partial^2 S}{\partial\phi(x)\partial\phi(y)}-\frac{1}{2}
\der{S}{\phi(x)}
\der{S}{\phi(y)}\right)+{\cal O}((\Delta t)^2)\right)
\label{SFP1}
\ee
The finiteness of the Langevin step induces
unavoidable corrections to the Boltzmann distribution. The Langevin
algorithm is not an exact one. These effects
can be reduced by applying a slightly more complicated Langevin
equation which cancels the ${\cal O}(\Delta t)$ corrections and is exact
up to ${\cal O} ({\left(\Delta t \right)}^2)$
\be
\phi(x,t_{n+1})=\phi(x,t_{n})
-\frac{\Delta t}{2} \sum_{y}\eps(x,y) \left(\der{S}{\phi(y,t_n)}+
\der{S(\bar{\phi})}{\phi(y,t_n)}\right)+\sqrt{\Delta t}\eta(x,t_n)
\label{mlang1}
\ee
\be
\bar{\phi}(x,t_{n})=\phi(x,t_{n})
-\Delta t \sum_{y}\eps(x,y) \der{S}{\phi(y,t_n)}+
\sqrt{\Delta t}\eta(x,t_n)
\label{mlang2}
\ee
But in practice there is no real gain in doing so.
A more interesting possibility
is to make use of the hybrid algorithm\cite{hyb}. This possibility
in the context of random surfaces will be explored elsewhere.

To gain some insight on the size of the deviations of the
equilibrium action with respect to the bare action one starts
with  it is useful to consider the effective action proposed
in \cite{Wh1}, which  reproduces  many features of the model.
In this case we have a $p$-dependent renormalization of our
parameters. For instance if $m$ is big enough the rigidity
coupling changes as
\be \kappa \to \kappa(1+\frac{\pi^2\Delta t}{4}b) \ee
where $b$ is a parameter which can be estimated and depends on $m$ and
$\kappa$. We shall use Langevin steps in the region $\Delta t=10^{-5}$
to $\Delta t=10^{-4}$ and we conclude that
the renormalization of the parameters due to the lack of exactness
of the algorithm is of the order of $0.1\%$.
We will see this explicitly when we obtain some observables.

\section{The Praxis}

In the previous section we described our method in very
general terms. When the method confronts reality a number of
technical issues have to be dealt with. We will set our
simulation of crystalline random surfaces on a
regular lattice
with the topology of a torus.
One of the advantages of FALA is the fact that the code is
easily vectorized and parallelized. FALA involves the inversion
of a matrix which with the help of fast Fourier transforms
can be reduced to a manageable $N\log N$ computer cost, but only
if our linear sizes are powers of 2.

When embedded in ${\bf R}^3$
each vertex has coordinates $x_i$ and each triangle has a
normal $n_I$. The action contains two pieces: an `area' term and a
rigidity (extrinsic curvature) one.
 After scaling the surface tension to 1
it reads
\be
S=\frac{1}{2}\sum_{i,j}{\left(x_i -x_j\right)}^2+\kappa
\sum_{I,J}\left(1- n_{I}\cdot n_{J}\right)\equiv A+\kappa S_{EC}
\label{action}
\ee
(From now on we will stick to the following notation: sites and
triangles are
labelled with lowercase and uppercase indices, respectively).
The partition function is
\be
Z=\int \prod_{i} d^3 x_i \delta^{(3)}(\sum_{i}x_i) \exp{(-S)}
\label{partition}
\ee

We must choose a precise form
for  $\eps(p)$ in \bref{eps1}. A physically motivated one would
be
\be
\eps(p)=\frac{\max_{p}\{\Delta \left(\Delta+m^2\right)\}}
{\Delta \left(\Delta+m^2 \right)},
\label{propg}
\ee
where $m$ is the inverse correlation length $\xi$ and $\Delta$ is
the lattice laplacian. The reason for this choice
is that the above expression
is ---up to a constant--- the exact two point function
in the large $D$ limit\cite{largeD}, which is described by
a gaussian theory. Gaussian models can be solved exactly
by using a FALA and the optimum function $\eps(p)$ turns out
to be the two point function. We shall use this form of
$\eps(p)$ for $D=3$, but the optimal value for $m$ will turn out not
to be $\xi^{-1}$.

The first point we have investigated is the need to incorporate a 2nd
order algorithm as the one described in equations
\bref{mlang1},\bref{mlang2}. When compared to the simplest first
order algorithm the computer time almost doubles.
This additional time would not be a problem if one could make the
Langevin step larger. As it is shown if fig. 1  the 2nd order algorithm
gives a better approximation to the actual value,
but to get acceptable  results the Langevin step has to
be of the same order of magnitude and no real gain in speed is obtained.

\begin{figure}
\centerline{\psfig{figure=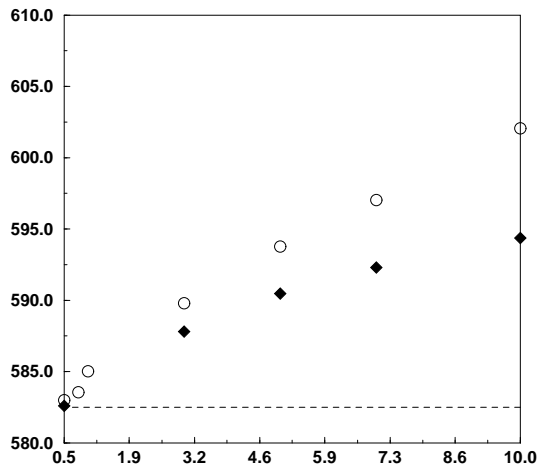,height=7cm}}
\caption{Comparison between first and second order Langevin algorithm.
The expectation value of the curvature is plotted versus the Langevin
step (in units of $10^{-4}$) for  $N=256,\kappa=0.5$.
The circles indicate the results obtained with a first order algorithm,
filled diamonds with a second order one and the dashed lines are the
exact result one would obtain with an exact algorithm (like Metropolis)
by performing the simulation at the actual coupling.}
\label{fig:1}
\end{figure}

Once this issue is settled, one notices immediately that
the algorithm does not
always reproduces the known results (some expectation values
can be computed exactly, see section 5).
The situation worsens for large
Langevin steps and for large values of $\kappa$, but even for very
small values of $\Delta t$ an analysis of the third
dispersion of the area clearly shows abnormally large values. After
dividing the data in small bins we can see the origin of the
difficulties in fig. 2. From time to time a
huge fluctuation takes place which spoils the statistical sample.
Needless to say that a second order algorithm
(or indeed any algorithm of any finite order) does not solve
the problem. What is causing these fluctuations?

\begin{figure}
\centerline{\psfig{figure=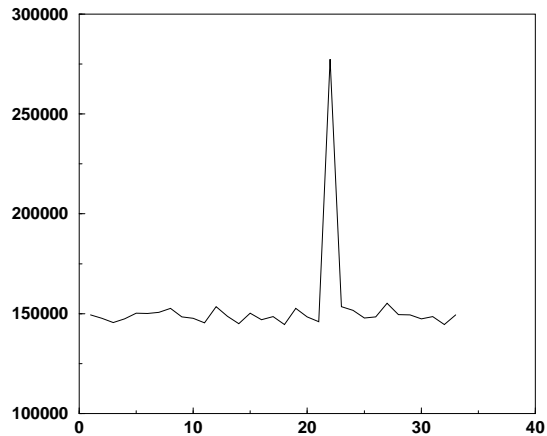,height=7cm}}
\caption{ Average value of $A^2$ per bin as a function of the bin
number. At bin number 22 there is a huge fluctuation.}
\label{fig:2}
\end{figure}

It should be clearly understood that the huge fluctuations
are an artifact of the algorithm requiring a finite $\Delta t$ and
 thus unphysical.
To see this we can introduce a parameter
$r$. At each step we measure the relative change
of the observables;
if it is greater than $r$, the whole configuration is
rejected, we generate a new set of random numbers and go on.
Typically one obtains that in $10^5$ configurations and $r=0.1$ just 5
or 6 configurations are rejected. With such small number of rejected
configurations it is clear that the convergence to the Boltzmann
configuration is not spoiled, but it is unclear whether all
configurations that are rejected are truly `unphysical'.
One also notices that while the number of
such huge fluctuations is essentially independent of $\Delta t$,
their magnitude does increase with the size of the Langevin step.
The results are presented in Table 1.

\begin{table}
\begin{center}
\begin{tabular}{|c|l|l|l|}                                \hline
$Observable   $  &$r=\infty$ &$r=0.1    $ &$exact$   \\ \hline
$ A           $  &$387.7   $  &$381.1     $ &$382.5$   \\ \hline
$ \Delta A    $  &$758.1   $  &$406.7     $ &$382.5$   \\ \hline
$ \Delta^2 A  $  &${\cal O}(10^5) $  &$927.3     $ &$765.5$   \\ \hline
$ S_{EC}      $  &$324.7   $  &$316.3     $ &$312.1$ \\ \hline
\end{tabular}
\end{center}
\caption{Results with and without rejection.
$\Delta A$ and $\Delta^2 A$ are the first and second dispersion
of the area, respectively. The parameters of the run are: $ N=16\times
16,\kappa=0.70$. No errors are shown.}
\end{table}

If we examine the configuration immediately preceding the huge
fluctuation
we will see that it always contains a triangle of very small
area.
The derivative of the action in the extrinsic curvature term in the
action in \bref{action} has the form
\be
n_I\cdot\der{n_J}{x^{\alpha}_k}={1\over A_J} {\eps^{\beta \alpha \nu}
\left(n_I^{\beta}- n_I\cdot n_J n_J^{\beta} \right)
{\left(x_i-x_j \right)}^{\nu}},
\label{spike}
\ee
$I,J$ label neighbouring triangles, $i$,$j$,$k$ are the three vertices
of triangle $J$. $A_J$ is the area of the
triangle whose normal is $n_J$.
If in eq.\bref{spike} $A_J$ is the area that
shrinks to a small value the following update would blow up---except
if $\Delta t$ is tiny, in which case thermalization would be
problematic. In order to have a smooth integration of the
Langevin equation the parameter which must be small
at each step throughout the lattice
is not $\Delta t$ but rather
$\Delta t/A$. Now a smooth cut-off can be introduced
according to this intuition
\cite{Wh3}. Let us modify the extrinsic curvature term in \bref{action}
by
\be
n_I\cdot n_J\rightarrow  n_I\cdot n_J
\exp \left(-s \Delta t \left({1\over A_I}+{1\over A_J} \right) \right),
\label{modif}
\ee
where $s$ is a new dimensionless parameter. This change causes a
negligible additional cost in time and formally
does not spoil the convergence to the desired
Boltzmann distribution as only  the
${\cal O}(\Delta t)$ piece in \bref{SFP1} is affected. Of course
since $\Delta t\neq 0$ in practice the observables
are modified, although by fine tuning $s$ this modification can be
made very small. Fine tuning the $s$ parameter
is a real problem however as the optimal $s$ depends on $\Delta t$,
$\kappa$, $m$ and even on the volume of the system. Tables 2, 3  show
different values of $s$ and how they affect the observables
for different choices of the parameters of the simulation.
For the Langevin step that we will eventually use, $\Delta t=8\times
10^{-5}$ and for $m=1$ the optimal values for $s$  are: $s=10$ for
$N=16\times 16$, $s=5$ for $N=32\times 32$ and $s=2$ for
$N=64\times 64$, although small variations do not drastically alter
the results.
The introduction and ensuing fine tuning of $s$, although
certainly a nuisance, are not expected to influence at all
the critical behaviour on universality grounds; actually they may
even provide for additional leeway in getting close to the
fixed point.

\begin{table}
\begin{center}
\begin{tabular}{|c|l|l|}                                     \hline
$s     $ &$A            $  &$\Delta A       $            \\ \hline
$1     $ &$379.6        $  &$426.6          $            \\ \hline
$2     $ &$380.9        $  &$429.7          $            \\ \hline
$10    $ &$380.5        $  &$398.4          $            \\ \hline
\end{tabular}
\end{center}
\caption{Dependence on $s$. Parameters of the
run: $\kappa=0.80,
N=16 \times 16, m=1, N_{sweeps}=3 \times 10^5, \Delta t=6 \times 10^{-5}$
The exact values are: $A=\Delta A=382.5$. No errors
are shown.}
\end{table}

\begin{table}
\begin{center}
\begin{tabular}{|c|l|l|}                                      \hline

$s     $ &$A              $  &$\Delta A    $            \\ \hline
$0.1   $ &$1542.1         $  &$1788.1         $            \\ \hline
$0.3   $ &$1539.8         $  &$1577.6         $            \\ \hline
$1     $ &$1539.2         $  &$1539.9         $            \\ \hline
$10    $ &$1538.4         $  &$1498.3         $            \\ \hline
$100   $ &$1536.2         $  &$1485.4         $            \\ \hline
\end{tabular}
\end{center}
\caption{Dependence on $s$. Parameters of the run:
$\kappa=0.80, N=32 \times 32, m=1,N_{sweeps}=3
\times 10^5, \Delta t=8 \times 10^{-5}$
The exact values are: $A=\Delta A=1534.5$. No errors are shown.}
\end{table}

In principle we should expect that setting
 $m=1/\xi$ (the inverse of the correlation
length) would lead to autocorrelation times
growing logarithmically with the linear size of the
system (that is $z=0$). Using $m=\xi^{-1}$
 implies that near the critical surface very
small values for $m$ have to be used. As it was observed in
\cite{Wh1}
the algorithm is then somewhat unstable and a larger value
for $m$ had to be used in \cite{Wh1,Wh2,Wh3}.
After extensive testing we decided to use $m=1$ in formula \bref{propg}
in our FALA,
We have also checked that larger values of $m$ do not bring any
real improvement, provided that
$\Delta t$ and $s$ are properly fine tuned. One should bear this
in mind when comparing different Langevin simulations.

Let us now turn to the crucial issue of
autocorrelation times. For a given observable (say the area) we
consider
\be
C_A(t)=\frac{\langle A_t A_0 \rangle-\langle A_t \rangle \langle A_0
\rangle}{\langle A_0^2 \rangle -{\langle A_0 \rangle}^2}.
\ee
For large values of $t$, $C_A(t)$ behaves as
$ C_A(t)=\exp(-t/\tau_A)$,
$\tau_A$ being the autocorrelation time of the observable $A$.
The autocorrelation of the area turns out to be roughly independent
of the volume of the system, but all the other observables
autocorrelation times are certainly volume-dependent.
Of all the measured autocorrelation times
the one of $ x^2$ is systematically the
longest. The first thing one should worry is the dependence
of the autocorrelation time on $\Delta t$, $m$ and $s$.
For fixed $m$,
a relation of the form $\tau_A \sim {1/\Delta t}$ holds, while the
dependence on $s$ is negligible. Larger values of $m$ correspond to
larger autocorrelation times as expected from the
form of eq.\bref{lang2}.
Some feeling on all these dependences can be gathered from figures  3 and
4.

\begin{figure}
\centerline{\psfig{figure=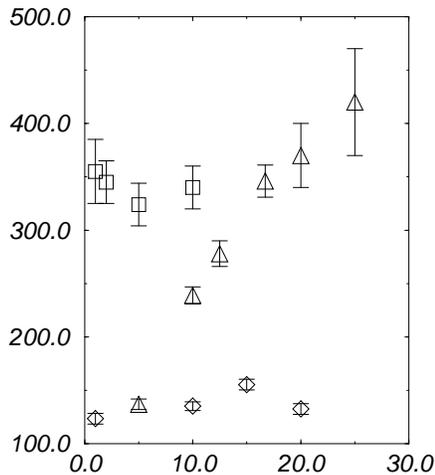,height=7cm}}
\caption{Autocorrelation times for $S_{EC}$ at
$\kappa=0.80$, $s=10$, $m=1$
as a function of $(\Delta t)^{-1}$ in units of $10^3$
(triangles). The squares show the dependence on $s$ of the
autocorrelation time for $\Delta t= 6\times 10^{-5}$,
 $\kappa=0.80$ and
$m=1$, and  correspond (left to right) to $s=1$,
2, 5  and 10. The diamonds
show the dependence on $s$ of the autocorrelation
 time for $\Delta t=8\times
10^{-5}$, $\kappa=0.80$ and $m=0.125$, and
correspond (left to right) to
$s=1$, 10, 15, 20.
In all cases the size is $16^2$ and the
statistics $5\times 10^5$ sweeps.}
\label{fig:3}
\end{figure}

\begin{figure}
\centerline{\psfig{figure=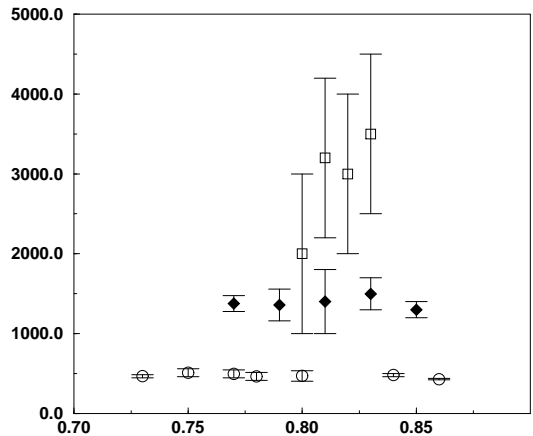,height=7cm}}
\caption{Autocorrelation times for $x^2$ as a function of $\kappa$
for different volume sizes: circles $16 \times 16$, filled diamonds
$32 \times 32$, squares $64 \times 64$. In the latter case  error
bars are big due to the lack of enough statistics}
\label{fig:4}
\end{figure}

In the light of the previous discussion it is clear that the
most relevant decision is to choose a sound value for $\Delta t$.
The best value is obtained by balancing systematic and statistical
errors. $\Delta t$ must be small enough so that
the equilibrium action is close enough to the bare action.  Yet
$\Delta t$ has to be large enough so that we get reasonably short
autocorrelation times and thus manage to sample configuration space
efficiently. The systematic error will be $\sigma_{syst}=a \Delta t$,
where $a$ is a constant whose magnitude we can determine
from the effective action proposed in \cite{Wh1}
 (see section 2) or simply
estimate from our data.
On the other hand, the statistical error will approximately be
$\sigma_{stat}=\sqrt{\tau/N}$, $\tau$ being a typical autocorrelation
time and $N$ the total number of sweeps that we can perform.
We already know that $\tau= b (\Delta t)^{-1}$ (for fixed
$\kappa$ the proportionality constant depends on $m$ and for $m=1$
is of ${\cal O}(10^{-2})$. Adding both errors in quadrature we get
that the optimal $\Delta t$ is
\be
\Delta t \sim (b/a^2 N)^{1/3}
\ee
This yields Langevin steps of ${\cal O}(10^{-3})$. This argument
is of course only indicative and further analysis shows that
it is actually better to use a slightly smaller step, of
${\cal O}(10^{-4})$, the reason being that due to the
difficulties that led to the introduction of the parameter $s$
the systematic errors are actually underestimated in the above argument.

\section{The Crumpling Transition}

The transition separating smooth from crumpled surfaces
 has attracted the
attention of workers in string theory and condensed matter alike.
There exist in the literature a variety of numerical and
analytical studies. At this point is where the usefulness of a
MCRG calculation becomes apparent. In the usual numerical
studies only the exponent $\nu$ (and those related to it by
hyperscaling relations) have been computed. In the analytical
analysis only the exponents related to the Hausdorff dimension have
been computed so far using a variety of approximations.
MCRG gives first-principles estimates for both exponents in a
single simulation.

Let us briefly review the definitions of the critical indices.
The specific heat is defined as
\be
C_v=\frac{\kappa^2}{N}\left(\langle S_{EC}^2 \rangle-
{\langle S_{EC} \rangle}^2 \right).
\ee
In the scaling region, where the correlation length $\xi$ is
large the exponents $\alpha$
and $\nu$ are defined by
\be
C_v \sim {\left(\kappa_{c}-\kappa\right)}^{-\alpha},
\qquad
\xi \sim {\left(\kappa_{c}-\kappa\right)}^{-\nu}.\ee
In \cite{Wh3} the well known hyperscaling relation $\alpha=2-\nu d$
has been tested numerically (within errors).

The Hausdorff or fractal dimension is defined by
\be \langle x^2 \rangle \sim N^{2/d_H}. \label{had}\ee
It is related to the $x$-field anomalous dimension
 $\eta$ \cite{am1},\cite{dav1}
defined by
\be \langle x(\xi_1) x(\xi_2) \rangle \sim
\frac{1}{{\|\xi_1-\xi_2\|}^\eta},
\ee
by the relation valid near $\kappa_c$
\be d_H=-4/\eta .\ee
(Notice that this definition of $\eta$ is not the one used in
\cite{am1}.)
We remark that $\eta$ must be a negative quantity, but from now on we
will omit the negative sign in $\eta$.

It is common in models of $2d$ gravity to define a mass gap
as the decay of the puncture-puncture correlation function. As the
mass gap vanishes  a critical index $\nu_{string}$
 is defined. This $\nu_{string}$ is
not related to $\nu$, but rather to the
Hausdorff dimension $d_H$ by $d_H=1/\nu_{string}$.

The first numerical studies \cite{oldsim,am1} established
the order of the transition. In \cite{am1} some estimates for the
Hausdorff dimension were presented. They reported $d_H=4.0\pm 0.4$,
although error bars were probably underestimated. In \cite{oldsim}
the tentative estimate $d_H\simeq 3$ was presented but errors were not
assessed.
Many authors have obtained estimates for the
critical exponents $\nu$ and $\alpha$. We summarize them in Table 4.
We regard the results in \cite{Wh3}
as the most reliable ones because they were obtained with bigger
lattices and show a reasonable agreement between finite size analysis
and a direct fit in the
scaling region, while previous simulations
\cite{Wh1},\cite{Wh2} and \cite{nos} were not entirely consistent.
Obviously there
is a lot of room for improvement in determining these exponents which
should identify the relevant CFT at the transition.

\begin{table}
\begin{center}
\begin{tabular}{|c|l|l|l|}                                      \hline
Authors         &$\alpha  $  &$\nu     $&$\alpha(finite size)$\\\hline
\cite{Renken}   &$        $  &$0.78(3) $&$0.44(5) $           \\ \hline
\cite{nos}      &$0.24(15)$  &$0.89(8) $&$0.70(50)$           \\ \hline
\cite{Petersson} &$        $  &$0.60(20) $&$0.71(5) $           \\ \hline
\cite{Wh3}      &$0.58(10)$  &$0.71(5) $&$0.54(10)$           \\ \hline
\cite{falcioni} &$        $  &$        $&$0.5     $           \\ \hline
\end{tabular}
\end{center}
\caption{Current estimates of the critical indices associated to
the crumpling transition}
\end{table}

The first analytical studies were focused on the large $D$ properties
\cite{largeD}, $D$ being the dimensionality of the ambient space.
At $D=\infty$ the $\beta$ function can be computed
exactly and there are no other zeroes than the trivial ones.
In  \cite{dav1} a model very similar to the one we are considering
 was studied. The $\beta$
function was computed in the large $D$ limit including the first
subleading term. Another fixed point was found and the Hausdorff
dimension obtained, with the result\cite{dav1}
$ d_H=2D/(D-1)$. At
$D=3$, $d_H=3$.
In \cite{Leo} an $\eps$ expansion was performed for a
variety of surface
models. The relevant result for us is
$ d_H=2.73 $,
which is compatible with the previous one up to ${1/D}$ corrections.
We are not aware of any analytical estimates for
$\nu$.

\section{The Results}

In our renormalization group analysis we have considered
9 different operators. They are lattice transcriptions of
simplified versions of
the continuum operators that would appear in a derivative
expansion. Their actual form is given in the Appendix.
There is a large freedom in choosing the precise discretized
form of a continuum operator, but the ambiguities associated
to this freedom should correspond to irrelevant operators,
representing non-universal corrections which vanish as positive
powers of the lattice spacing modulo logarithms
when the continuum limit is approached.
Provided that the ones we include form a reasonably complete set
with all the right symmetries up to a given dimension we should be
on safe ground. We shall, at any rate, evaluate the errors
associated to the truncation of the series of operators and to
discretization ambiguities.

After elimination of the higher $p$-modes we Fourier transform
back to position space and end up with a lattice with half the
number of points in each direction. The rescaling factor
discussed in section 2 will be determined by demanding that
near a fixed point the expectation value of any bare and
renormalized operators should coincide. We therefore allow $\eta$
to vary freely and choose the value that best fits this
requirement.
Only if we
are close enough to the fixed point the rescaling of the fields
has a universal meaning and is the conformal anomalous dimension.
Some operators (those constructed with
normals alone) contain no $x$-fields and therefore their
matching will tell us in a unambiguous way
how close we are to a fixed point.
We have found that
one blocking is not enough to bring us near
the fixed point when we start with our bare original action, no matter
how close to the
critical surface we position ourselves. A second blocking is required
to take us to the vicinity of the fixed point.
Then the results are very satisfactory.

Let us consider the rescaling $ x \rightarrow
\lambda x $ in the bare action
 \bref{action}. Using this rescaling we can derive
the following relations
\be   \langle A \rangle=\frac{(N-1)D}{2}, \label{sca1}\qquad
\langle A^2 \rangle-{\langle A \rangle}^2 =
\frac{(N-1)D}{2},
\label{sca2}
\ee
\be
\langle A^3 \rangle-3\langle A^2 \rangle{\langle A \rangle}^2
+2{\langle A \rangle}^3=(N-1)D.
\label{sca3}
\ee
More relations, involving higher powers of $A$ may be obtained along
the same lines, but they are of no use to us.
These relations assume that the surface tension is set to unity,
but it can be easily restored by dimensional analysis.
To derive formulae \bref{sca2} and \bref{sca3}
we have only assumed that the extrinsic curvature part of the action is
scale invariant. If by any chance,
after a renormalization group transformation,
only operators involving normals are generated
the above formulae should hold exactly.
There is really no reason why higher dimensional
operators containing powers
of $x$ should not appear, but we can have an idea whether they
trigger violations of the above scaling relations
by checking the validity of
\bref{sca2},\bref{sca3} after a renormalization group blocking.

At volume $32 \times 32$ and $\kappa=0.82$ our simulation gives
per site
\be
\langle A\rangle =1.503  \qquad \langle A^2\rangle =2.260
  \qquad \langle A^3\rangle =3.400
\ee
These results agree with those obtained from \bref{sca2} and
\bref{sca3} after a tiny rescaling of $\mu$ (of ${\cal O}(10^{-3})$),
showing
that the surface tension does not get a large modification
due to the use of the (non-exact) Langevin algorithm,
in accordance with our
expectations discussed in section 2.
After blocking down to $16\times 16$ and choosing $\eta$
such that the same value of the area per site is reproduced
we have
\be
\langle A\rangle=1.503 \qquad \langle A^2\rangle=2.260
 \qquad \langle A^3\rangle =3.404
\ee
The agreement, which is not peculiar to this value of $\kappa$ or to this
volume, is impressive and it suggests that the effect of scale
non-invariant operators is not very important.

We shall order the operators in
the $T$ matrix in the following way: (a) operators  made out of
normals only, (b) operators that contain both normals and
coordinates $x_i$ and (c) those which
contain no normals at all. Within each group we shall order them
by their dimensionality under the rescaling $x\to\lambda x$.

As we will see below the operators
\be
\Delta x \Delta x, \qquad (\nabla x)^4
\label{badguys}
\ee
induce imaginary parts in the largest eigenvalue when they
are included in the renormalization group analysis. We do not
understand  this fact very well. There is no reason to forbid these
operators in our basis. In fact
it was found in \cite{Wh1} that the large distance
properties of crystalline random surfaces were well
described by a term like the first one in eq. \bref{badguys}. This
operator is nothing but another
representation of the same operator as the extrinsic
curvature itself (in fact some early simulations used this operator
instead of $S_{EC}$ to represent the rigidity term).
These operators are very sensitive to the actual value of $\eta^*$ one
uses and by choosing a smaller value for the anomalous dimension
the eigenvalues can be made real, but with large fluctuations. The
second eigenvalue is then always very close to one.
 Of course the anomalous
dimension must be gotten from the matching so this procedure is not
consistent and must be discarded.
(Incidentally, it is worth pointing out that the matching
for these operators is as good as for the rest).
 We are inclined to believe
that the difficulties with the imaginary parts could
 be solved by performing
further blockings,
as something similar happens with other operators when we compare
the first and second blockings.

Our strategy has been
the following. After finding the optimal values of $\Delta t$,
$m$ and $s$, we run with the action
\bref{action} for different values
of $\kappa$ until we get close to the critical surface
from both sides (smooth and crumpled phases).
Then we perform the following blockings: $64\times 64\to
32\times 32\to 16\times 16$, $32\times 32 \to 16\times 16\to
8\times 8$ and $16\times 16 \to 8\times 8$. We determine whether we
are in the vicinity of the
 fixed point and if so proceed to measure $\nu$ and $\eta$.
Additional runs have been used to
measure the autocorrelations.

Preliminary results of our work were presented in \cite{ET2}. Since
then we have accumulated many more configurations and
performed a second iteration of our
renormalization group transformation, which has proven essential
to pin down $\nu$ and, particularly, $\eta$. The results
differ to some extent from those presented in \cite{ET2} for
reasons that will be explained below.

Error bars have three different origins. The statistical error
is evaluated it by the usual binning procedures.
We have good control over the autocorrelation
times. In the case of a $64 \times 64$ system the largest
one (corresponding to $\langle x^2\rangle$) is in the range,
$2000 \sim 4000$ sweeps,
fortunately  by no means out of the reach of our computer facilities.
The second source of errors are more difficult to assess. They have to do
with the truncation of the number of operators entering the
renormalization, the precise determination of the fixed
point and, of course, $\Delta t$, errors as well as the
uncertainties related to the $s$ parameter. And, last but not least,
we have to deal with finite size errors. Critical exponents are affected
by errors due to the influence of the finiteness of the lattice on the
expectation values from which they are derived. However, what we call
systematic errors (i.e. the dispersion in eigenvalues or errors due
to truncation and discretization) are somewhat entangled with
genuine finite size errors. This probably means that our error bars are
probably overestimated.

All the numerical work was done with a CRAY-YMP, and SGI Power
Challenge L. A first renormalization
$32 \times 32 \rightarrow 16 \times 16$ with 500,000 configurations
takes about 1.5 hours of CPU time on a Cray-YMP.
Applying twice the renormalization group to a
 $64 \times 64$ system with a
similar number of configurations 
 in a SGI Power challenge L where three
processors were fully used by our code
takes about
52 hours of total CPU time or about 18 hours in real time.

\subsection{First Renormalization}

We shall not discuss in any detail the results of the
$16 \times 16\to
8 \times 8$ blocking. It is only important to remark that
many of the general trends are already present in such a small
system. The value obtained for the anomalous dimension
is $\eta=1.05 (40)$.
All errors quoted in this
subsection merely reflect the systematic uncertainties due to the
dispersion in values in the different operators
and do not include statistical errors. The errors
due to finite-size effects are not estimated here either.
This is sufficient for our purposes.

Let us apply the MCRG to a system of size
$32 \times 32$ to get a $16 \times 16$ lattice. We will
present results for $\kappa=0.805$ (crumpled phase) and
$\kappa=0.82$ (smooth phase) with
$2\times 10^6$ sweeps each, although we have run for $\kappa$ values
between 0.79 and 0.83.
For $\kappa=0.805$
a correlation length of about 10 lattice units was reported to us
\cite{personal}, so finite size effects,
though certainly present, are small.
Unfortunately the blocking $32\times 32\to 16\times 16$ does not bring us
close enough to the fixed point. (Recall that in $\lambda \phi^4_3$ one
MCRG iteration was enough to get an excellent matching; obviously for
rigid surfaces the fixed point lies far away from the canonical surface.)
This is evidenced by the results presented in Tables 5, 6,
where we estimate the anomalous dimension by attempting to match
the bare and renormalized operators. Each operator requires a different
$\eta$, with a large dispersion for the values,
a clear signal that we are not yet in the vicinity
of the fixed point, where an universal $\eta$ would suffice. Notice that
the matching is not good for the operators $O_1$ to $O_4$ either, which
depend only on normals and thus are $\eta$ independent.

\begin{table}
\begin{center}
\begin{tabular}{|c|l|l|l|}                                     \hline
	 &$32\times 32$&$16\times 16 (\eta)$&$16\times 16$\\ \hline

$ A    $&$ 1.502   $&$1.502 (1.097)     $&$1.462            $\\ \hline
$ A^2  $&$ 2.259   $&$2.260 (1.097)     $&$2.141            $\\ \hline
$ A^3  $&$ 3.397   $&$3.404 (1.097)     $&$3.139            $\\ \hline
$S_{EC}^2$&$ 1.720   $&$1.400             $&$1.400            $\\ \hline
$ O_1  $&$ 1.310   $&$1.176             $&$1.176            $\\ \hline
$ O_2  $&$ 1.373   $&$1.210             $&$1.210            $\\ \hline
$ O_3  $&$ 5.714   $&$8.100             $&$8.100            $\\ \hline
$ O_4  $&$ 0.885   $&$0.760             $&$0.760            $\\ \hline
$ O_5  $&$ 2.024   $&$2.020 (1.510)     $&$2.617            $\\ \hline
$ O_6  $&$ 1.998   $&$2.040 (1.55)     $&$2.712            $\\ \hline
$ O_7  $&$ 4.871   $&$4.84(0.97)       $&$3.725            $\\ \hline
$ O_8  $&$ 4.465   $&$4.33(0.7)        $&$3.210            $\\ \hline
\end{tabular}
\end{center}
\caption{Matching between big and small lattices
for $\kappa=0.805, s=5, m=1$. If we allow $\eta$ to vary independently
for each operator and impose matching
we obtain the values in the second column (the corresponding $\eta$ is
shown in brackets).
If instead we use the average $\eta$, which we call
$\eta^*$, we obtain the values in the third column. The expectation
values of the operators in this and in similar tables are given
per site.}
\end{table}

\begin{table}
\begin{center}
\begin{tabular}{|c|l|l|l|}                                      \hline
	&$32\times32$&$16\times16 (\eta) $&$16\times16$     \\ \hline
$ A     $&$ 1.503   $&$1.502 (1.124)   $&$1.475            $\\ \hline
$ A^2   $&$ 2.260   $&$2.260 (1.124)   $&$2.180            $\\ \hline
$ A^3   $&$ 3.400   $&$3.404 (1.124)   $&$3.225            $\\ \hline
$ S_{EC}^2 $&$ 1.597   $&$1.256        $&$1.256            $\\ \hline
$ O_1   $&$ 1.262   $&$1.113           $&$1.113            $\\ \hline
$ O_2   $&$ 1.295   $&$1.114           $&$1.114            $\\ \hline
$ O_3   $&$ 5.404   $&$7.420           $&$7.420            $\\ \hline
$ O_4   $&$ 0.848   $&$0.711           $&$0.711            $\\ \hline
$ O_5   $&$ 1.997   $&$2.000 (1.524)   $&$2.591            $\\ \hline
$ O_6   $&$ 2.006   $&$2.04  (1.60)   $&$2.782            $\\ \hline
$ O_7   $&$ 4.864   $&$4.88  (0.96)   $&$3.762            $\\ \hline
$ O_8   $&$ 4.374   $&$4.5(0.6)       $&$3.104            $\\ \hline
\end{tabular}
\end{center}
\caption{Same as in the previous table, but for $\kappa=0.820$}
\end{table}

In order to obtain $\nu$ and before proceeding to a second iteration of the
renormalization group transformation we need to rescale the fields
$x$ by choosing some prescription. The natural one is to demand that
after a renormalization group transformation the normalization of the
area term in the action is preserved.
The fact that violations to the relations \bref{sca2} and \bref{sca3}
are extremely small after a blocking suggests that, with very good
accuracy, the blocked action has still the same invariance $x\to \lambda
x$, $\mu\to \lambda^{-2}\mu$ that the original action had. Assuming that
this symmetry is exact the absolute normalization of the area term in the
blocked action can
be obtained by comparing $\langle A\rangle$ and, say, $\langle A^2\rangle$,
and the rescaling of the $x$-fields computed.
If we were at the fixed point this rescaling would suffice to
match all operators and would give to us the universal anomalous dimension
$\eta$.
When this is not quite the case, as it happens here,
we simply use the rescaling
obtained from the area and move on
to the next blocking until we get close
enough to a fixed point where a universal value for $\eta$ exists.
Because in practice all observables are affected by errors it
seems a good
idea to use, to a limited extent, the rescalings which are suggested
by operators other than the area too. We adopt a compromise solution
were the final value of the rescaling that is used to derive $\nu$
is determined by performing a weighted average of the different values
of $\nu$ in which
area-type operators weigh more. Then we obtain
$\eta^{*}=1.14(5)$ and
$\eta^{*}=1.15(5)$ for $\kappa=0.805$ and $\kappa=0.820$, respectively.
The $\chi^2/$d.o.f.
is in both cases of ${\cal O}(100)$, which just reflects we are
not quite at the fixed point.
Using the values for $\eta^*$ obtained in this way we can determine the
eigenvalues of the matrix $T$
and find $\nu=1.25(50)$. In short,
$\eta^*$ cannot be interpreted
as the anomalous dimension; it is just a convenient way of parametrizing
the rescaling of the fields.

In the preliminary results presented in \cite{ET2} only one blocking was
being performed and the rescaling of the area was used to define
the anomalous dimension. Although this is certainly a proper
way of getting the rescaling of the fields,
the resulting $\eta^*$ cannot be identified
with the anomalous dimension because we were too far from the fixed point.
We did, in fact, misjudged the distance to the fixed point. The second
blocking results clarify this issue completely.

\begin{table}
\begin{center}
\begin{tabular}{|c|l|l|}                                      \hline
$ \#   $ &$\kappa=0.805$           &$\kappa=0.820  $           \\ \hline
$  1   $ &$1.60          $  &$1.58          $            \\ \hline
$  2   $ &$1.54          $  &$1.53          $            \\ \hline
$  3   $ &$1.86          $  &$1.86         $            \\ \hline
$  4   $ &$2.04          $  &$2.24          $            \\ \hline
$  5   $ &$1.99          $  &$2.17          $            \\ \hline
$  6   $ &$1.80          $  &$1.98           $            \\ \hline
$  7   $ &$2.47*          $  &$2.59*          $            \\ \hline
$  8   $ &$2.46*          $  &$2.59*           $            \\ \hline
\end{tabular}
\end{center}
\caption{
First blocking in the
$32\times 32\to 16\times 16 \to 8\times 8$ sequence.
Evolution of the largest eigenvalue of the matrix $T$ as
more and more operators are included in the MCRG.
The asterisk
indicates that the eigenvalue develops an imaginary part.}
\end{table}

We can also apply one iteration of the MCRG to a
system of $64 \times 64$ down to $32\times 32$.
For this size, autocorrelation for observables are relatively
large (see fig. 4) and better statistics are required. Some
$10^7$ configurations have been accumulated.
We just present here the results for
$\kappa=0.82$:
$\eta^*=1.22(30)$, $\nu= 1.06(30)$,
with a $\chi^2/$d.o.f$\sim 10^2$.
There is no real gain in precision
 and the results are not better than
with smaller lattices. It is interesting to note that the `anomalous
dimension' increases with the volume.

\subsection{Second Renormalization}

From the results just discussed we conclude
that the canonical surface is
far from the fixed point. A second iteration of the renormalization
group transformation should get us much closer. Closer in any case
to the renormalized trajectory;
that is, the trajectory that goes in the direction of the
relevant operator of the theory near this fixed point. (We will later
see that there is only one relevant direction.)

We shall apply the by now familiar machinery to the two-step blockings
$32\times 32\to 16\times 16 \to 8\times 8$ and $64\times 64\to 32\times
32\to 16\times 16$.  Both the determination of $\eta^*$ and $\nu$ in the
second iteration are independent of the value of $\eta^*$ used in the
first one, although the precise numerical values of the different
observables is sensitive to the first rescaling.  In order to present
our data we will take as the anomalous dimension of the first
renormalization the one obtained from equations \bref{sca2},\bref{sca3}
for the reasons discussed in the previous subsection.  After following
the usual strategy we see in Tables 8, 9 the results of the second
blocking.  It is obvious that there is a tremendous improving with
respect to the results presented in Tables 5, 6. We no longer have a
large dispersion in the $\eta$ exponent and the matching is now much
better.  For operators that are made only of normals the agreement is at
the $20\%$ level.  Relations such as \bref{sca2},\bref{sca3} are
satisfied with very good accuracy.

If we perform a least square fit (now with equal weights for all operators)
we get for $\kappa=0.805$
$\eta^{*}=1.41(20)$ while
at $\kappa=0.820$ we get
$\eta^{*}=1.44(20)$.
In both cases $\chi^2/$d.o.f$\simeq 50$.
The Hausdorff dimension is $d_H=2.83(23)$.
In spite of the still somewhat
big error bars
we can already
compare with the analytical estimates of section 4 with good agreement.

\begin{table}
\begin{center}
\begin{tabular}{|c|l|l|l|}                                     \hline
	&$16\times16$&$ 8\times8 (\eta) $&$ 8\times8       $\\ \hline
$ A     $&$ 1.502   $&$1.500 (1.400)     $&$1.491            $\\ \hline
$ A^2   $&$ 2.260   $&$2.261 (1.400)     $&$2.231            $\\ \hline
$ A^3   $&$ 3.400   $&$3.415 (1.400)     $&$3.365            $\\ \hline
$S_{EC}^2$&$ 1.400   $&$2.350            $&$2.350            $\\ \hline
$ O_1   $&$ 1.176   $&$1.522           $&$1.522            $\\ \hline
$ O_2   $&$ 1.210   $&$1.177           $&$1.177            $\\ \hline
$ O_3   $&$ 8.101   $&$12.44           $&$12.44            $\\ \hline
$ O_4   $&$ 0.760   $&$1.000           $&$1.000            $\\ \hline
$ O_5   $&$ 2.689   $&$2.690 (1.490)   $&$2.750            $\\ \hline
$ O_6   $&$ 2.750   $&$2.700 (1.34)   $&$2.534            $\\ \hline
$ O_7   $&$ 3.931   $&$3.931 (1.378)   $&$3.720            $\\ \hline
$ O_8   $&$ 3.300   $&$3.34  (1.6)   $&$3.730            $\\ \hline
\end{tabular}
\end{center}
\caption{ Matching between big and small lattices after the second
iteration of the MCRG on a $32\times 32 $ system. $\kappa=0.805,s=5,m=1$}
\end{table}

\begin{table}
\begin{center}
\begin{tabular}{|c|l|l|l|}                                      \hline
	&$16\times16$&$ 8\times8 (\eta) $&$ 8\times 8      $\\ \hline
$ A     $&$ 1.502   $&$1.500 (1.433)     $&$1.491            $\\ \hline
$ A^2   $&$ 2.260   $&$2.260 (1.433)     $&$2.231            $\\ \hline
$ A^3   $&$ 3.404   $&$3.420 (1.433)     $&$3.365            $\\ \hline
$ S_{EC}^2$&$ 1.260   $&$2.174           $&$2.174            $\\ \hline
$ O_1   $&$ 1.113   $&$1.459           $&$1.459            $\\ \hline
$ O_2   $&$ 1.115   $&$1.670           $&$1.670            $\\ \hline
$ O_3   $&$ 7.420   $&$11.73           $&$11.73            $\\ \hline
$ O_4   $&$ 0.713   $&$0.960           $&$0.960            $\\ \hline
$ O_5   $&$ 2.640   $&$2.635 (1.50)     $&$2.750            $\\ \hline
$ O_6   $&$ 2.833   $&$2.800 (1.30)     $&$2.534            $\\ \hline
$ O_7   $&$ 3.900   $&$3.901 (1.405)     $&$3.720            $\\ \hline
$ O_8   $&$ 3.161   $&$3.3(1.6)         $&$3.730            $\\ \hline
\end{tabular}
\end{center}
\caption{ Matching between big and small lattices after the second
iteration of the MCRG on a $32\times 32 $ system. $\kappa=0.820,s=5,m=1$}
\end{table}

Let us now compare the values of$\langle x^2\rangle$ in the big
and small lattices
\be
\langle x^2\rangle_{16 \times 16}=4.84, \qquad
\langle x^2\rangle_{8 \times 8}=1.78.
\ee
Using (see previous section)
\be
d_H=\frac{4\log2}
{\log{\langle x^2\rangle_N}-\log{\langle x^2\rangle_{\frac{N}{2^d}}}}
\label{logdh}
\ee
we obtain $d_H=2.77(10)$ in good agreement with the one already
obtained. We can even check the consistency by employing higher
cumulants of $x^2$ and still obtain the same result.

With the above values for $\eta^*$ we get the series of eigenvalues
displayed in Table 10.
The eigenvalues are all bigger than 2, implying a second order
transition. We obtain
$\nu=0.87(10)$.
Error bars just reflect the dispersion of the eigenvalues
of the $T$ matrix; no statistical or finite-size
errors are considered yet.

\begin{table}
\begin{center}
\begin{tabular}{|c|l|l|}                                      \hline
$ \#   $ &$\kappa=0.805 $    &$\kappa=0.820 $              \\ \hline
$  1   $ &$2.08           $  &$2.04           $            \\ \hline
$  2   $ &$2.10           $  &$2.05           $            \\ \hline
$  3   $ &$2.34           $  &$2.24           $            \\ \hline
$  4   $ &$2.30           $  &$2.21           $            \\ \hline
$  5   $ &$2.32           $  &$2.36           $            \\ \hline
$  6   $ &$2.28           $  &$2.33           $            \\ \hline
$  7   $ &$1.99*          $  &$2.09*          $            \\ \hline
$  8   $ &$1.91*          $  &$2.00*          $            \\ \hline
\end{tabular}
\end{center}
\caption{
Second blocking in the $32\times
32\to 16\times 16\to 8\times 8$ sequence.
Evolution of the largest eigenvalue of the matrix $T$ as an
increasing number of operators are included in the MCRG.}
\end{table}

The above results are very encouraging, but not fully satisfactory
yet. It is clear that the matching is much better, implying that
we are in the vicinity of the fixed point. Perhaps it should not
be surprising that the dispersion in $\eta$ is still large because
after the second application of the MCRG
to a $32\times 32$ lattice we end up with a $8 \times 8$ lattice.
In principle we should not have any more
difficulties on a $8\times 8$ lattice than we had in the
original $32\times 32$ one since
at the same time that the lattice is reduced the correlation length
is reduced by exactly the same amount.
We try to keep severe finite size  effects
under control by running at values of
$\kappa$ which are close enough to the
critical surface, yet with a correlation length smaller than our system
size. However
the `critical' behaviour for such small lattices shows a rather erratic
pattern\cite{nos} with
the critical region being significantly shifted.
In spite of this the results are reasonably good, showing that
the method is very robust.

It is clear that larger systems are called for. Let us
now turn to the
$64 \times 64 \rightarrow 32 \times 32\to 16\times 16$ sequence of
blockings.
While several values of $\kappa$
 were investigated in smaller lattices, due
to computer time availability  here
we will run only at
$\kappa=0.805$ and $\kappa=0.82$, collecting up to $10^7$
configurations.  The enlargement of 
the lattice size brings about a little
miracle. We see at once from
Tables 11, 12 that the matching between bare and
renormalized operators is now wonderful, particularly for $\kappa=0.820$. The
dispersion in the anomalous dimension reduces now
to a narrow width.
Note that operators that are
insensitive to the anomalous dimension show agreement which is in
some cases as good as a $1\%$. This is a matching of the same quality
as the one obtained in \cite{ET1} and it is the one we were after.

\begin{table}
\begin{center}
\begin{tabular}{|c|l|l|l|}                                     \hline
	&$32\times32$&$16\times16(\eta)$&$16\times16       $\\ \hline
$ A     $&$ 1.496   $&$1.496 (1.55)     $&$1.545            $\\ \hline
$ A^2   $&$ 2.238   $&$2.237 (1.55)     $&$2.401            $\\ \hline
$ A^3   $&$ 3.349   $&$3.3525(1.55)     $&$3.705            $\\ \hline
$S_{EC}^2$&$ 0.695   $&$0.740             $&$0.740          $\\ \hline
$ O_1  $&$ 0.830   $&$0.852           $&$0.852       $\\ \hline
$ O_2  $&$ 0.694   $&$0.760           $&$0.760            $\\ \hline
$ O_3  $&$ 4.431   $&$4.902           $&$4.902            $\\ \hline
$ O_4  $&$ 0.508   $&$0.510           $&$0.510            $\\ \hline
$ O_5  $&$ 2.451   $&$2.453 (1.51)    $&$2.467            $\\ \hline
$ O_6  $&$ 3.002   $&$3.010 (1.53)    $&$3.065            $\\ \hline
$ O_7  $&$ 3.763   $&$3.752 (1.50)    $&$3.741            $\\ \hline
$ O_8  $&$ 2.556   $&$2.51 (1.45)     $&$2.42            $\\ \hline
\end{tabular}
\end{center}
\caption{Matching between big and small lattices after
the second
iteration of the MCRG on a $64\times 64 $ system. $\kappa=0.805,s=2,m=1$}
\end{table}

\begin{table}
\begin{center}
\begin{tabular}{|c|l|l|l|}                                      \hline
	&$32\times32$&$16\times16 (\eta) $&$16\times16$\\ \hline
$ A     $&$ 1.503   $&$1.503 (1.585)     $&$1.565            $\\ \hline
$ A^2   $&$ 2.260   $&$2.260 (1.585)     $&$2.451            $\\ \hline
$ A^3   $&$ 3.397   $&$3.400 (1.585)     $&$3.843            $\\ \hline
$ S_{EC}^2$&$ 0.555   $&$0.553           $&$0.553            $\\ \hline
$ O_1   $&$ 0.744   $&$0.740           $&$0.740            $\\ \hline
$ O_2   $&$ 0.573   $&$0.606           $&$0.606            $\\ \hline
$ O_3   $&$ 3.617   $&$3.889           $&$3.889            $\\ \hline
$ O_4   $&$ 0.448   $&$0.435           $&$0.435            $\\ \hline
$ O_5   $&$ 2.382   $&$2.382 (1.525)     $&$2.380            $\\ \hline
$ O_6   $&$ 3.039   $&$3.034 (1.567)     $&$3.120            $\\ \hline
$ O_7   $&$ 3.757   $&$3.760 (1.527)     $&$3.768            $\\ \hline
$ O_8   $&$ 2.387   $&$2.36  (1.45)     $&$2.200            $\\ \hline
\end{tabular}
\end{center}
\caption{Matching between big and small lattices after the second
iteration of the MCRG on a $64\times 64$ system. $\kappa=0.820,s=2,m=1$.
The statistical error is $\pm 0.002$ in all observables.}
\end{table}

We notice that
for $\kappa=0.82$ we are particularly close
to the fixed point after two iterations, although $\kappa=0.805$ is also
good. We will draw our conclusions from the former
value of $\kappa$ were matching is slightly better. The
results are then
\be
\eta^{*}=1.527\pm 0.010\pm 0.006\pm 0.020
\label{etabar}
\ee
with a $\chi^2/$d.o.f$\simeq 1$. Errors due to truncation, statistics
and finite size effects are separately quoted (in this order; see the
discussion at the beginning of this section).
We remind the reader that to obtain the true anomalous dimension of the
field one should reverse the sign of $\eta^*$.
The Hausdorff dimension is
\be
d_H=2.62\pm 0.02 \pm 0.01 \pm 0.04
\ee
Where the first error comes from the dispersion of the anomalous
dimensions,
the second one is statistical and the third one an
estimate of the error due to finite size effects.
 A self consistent computation
of $d_H$ along the
lines of equation \bref{logdh},
 considering that from our simulation we get,
\be
\langle x^2\rangle_{32 \times 32}=20.98(8),\qquad
\langle x^2\rangle_{16\times16}=7.33(13),
\ee
yields
\be
d_H=2.64(5),\label{dh}
\ee
in good agreement with the previous estimate. Only the statistical error
is included.

A look at the table of eigenvalues shows that the dispersion has
reduced tremendously when compared to the previous cases. From the
results in Table 13
\be
\nu=0.85\pm 0.07\pm 0.02\pm 0.06 \label{nu}
\ee
The first error bar corresponds to the systematic
 uncertainty of the method,
the second one
corresponds to statistical errors, and the third one
estimates finite size errors, as before. Adding all errors
in quadrature we get $\nu=0.85\pm 0.09$.

\begin{table}
\begin{center}
\begin{tabular}{|c|l|}                                      \hline
$  \#   $ & $\lambda_h $                            \\ \hline
$  1   $ &$2.25(-)(2)     $             \\ \hline
$  2   $ &$2.23(-)(2)     $             \\ \hline
$  3   $ &$2.22(-)(2)     $             \\ \hline
$  4   $ &$2.15(-)(3)     $             \\ \hline
$  5   $ &$2.20(3)(5)     $             \\ \hline
$  6   $ &$2.18(3)(5)     $              \\ \hline
$  7   $ &$2.19(6)(6)*    $              \\ \hline
$  8   $ &$2.18(6)(6)*    $              \\ \hline
\end{tabular}
\end{center}
\caption{
Second blocking in the $64\times
64\to 32\times 32\to 16\times 16$ chain.
Evolution of the largest eigenvalue of the matrix $T$ as a function
of the number of operators included in the MCRG. The first error is due to
the uncertainty in the anomalous dimension $\eta^*$ (it does not affect
ops. 1-4), while the second one reflects the statistical error.}
\end{table}

In the $T$ matrix all the other eigenvalues are smaller than one so
there is only one relevant direction in the theory. We can give
a tentative value for the first subleading exponent $\omega$,
\be
\omega = -0.35^{+ 0.25}_{-0.50}.
\label{endo}
\ee
Equations \bref{etabar} to \bref{endo} are our final results.
If we compare our results with previous determinations
(see section 4) we see that they are in reasonable agreement.
$\nu$ is slightly higher than the most
recent estimates, with similar error bars. $d_H$ is slightly lower than
the theoretical predictions. The amount of
numerical work required to determine $\nu$ is
however substantially smaller than with other methods and,
 of course, we can
determine $d_H$ directly with a good assessment
 of the errors involved and,
perhaps more importantly, we gain a very good understanding of the
crumpling transition.  For instance, it is very
 interesting to see how the
renormalization group brings together after
 two blockings the trajectories
that started at both sides of the critical surface.

\subsection{$\beta$ Function}

MCRG has permitted us to compute the critical
exponents but we
 have said very little about the flow of the renormalization
group and the effective action one obtains after integrating high
$p$-modes. Within our framework it is not possible to compute
the renormalized couplings so these important problems would require a
separated study.
However, something can be said about the
general form of the $\beta$ function.
There are two realistic scenarios
as shown in figure [5]. Our results give strong support
that case B is the one that is realized, even if the
evidence is not fully conclusive.

Let us consider the quantity $R_{rg}$, defined as
\be  R_{rg}=\frac{\langle S_{EC}\rangle_N}{2^d\langle
S_{EC}\rangle_{\frac{N}{2^d}}},
\ee
That is $R_{rg}$ is the ratio of the average of extrinsic curvatures
before and after renormalization with the convenient factor that
makes it an intensive quantity. Now,
if $R_{rg} > 1$ then the surface is less crumpled in the infrared whereas
if $R_{rg} < 1$ it becomes more crumpled as we move to larger distances.
On a
$64 \times 64$ lattice at $\kappa=0.83$  we can compute the
average extrinsic curvature and obtain
\be
    \langle S_{EC}\rangle_{64 \times 64}=4829(2), \qquad
    \langle S_{EC}\rangle_{32 \times 32}=857(1).
\label{curv}
\ee
Then at $\kappa=0.83$ (smooth phase) we have $R_{rg}=1.50$, clearly the
surface is less crumpled
in the infrared, in agreement with case B.

\begin{figure}
\centerline{\psfig{figure=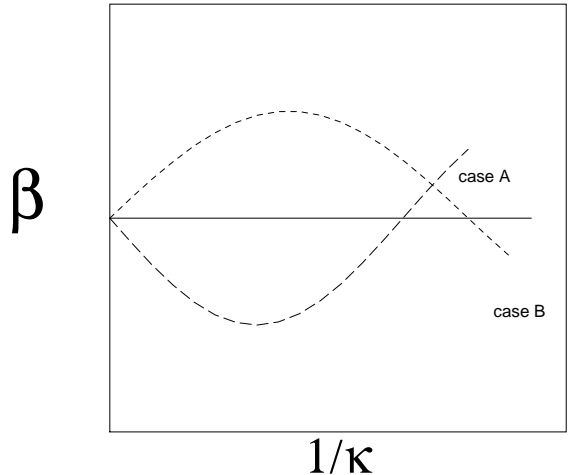,height=7cm}}
\caption{$\beta$ as a function of $1/\kappa$ in arbitrary units. The two
relevant cases for us are depicted, In case A the origin is an
ultraviolet fixed point while in case B it is infrared stable.}
\label{fig:5}
\end{figure}

We can see this behaviour in another way. We apply MCRG at $\kappa=0.83$
and we end up with a smaller lattice. Let us make an estimate of the value
of $\kappa_{eff}$ which we would have to use in the small system to
obtain the value quoted in \bref{curv}. A look at a simulation in a
system $32 \times 32$\cite{Renken} shows that
$\kappa_{eff} > 0.9$ so the flow goes according to case B. The analysis
for other $\kappa$ values is also as clean in giving evidence for
the same shape of the $\beta$-function.

Unfortunately in the crumpled phase things are not so clear. If we
work with a system at $\kappa=0.79$ we obtain that $R_{rg}=1.05(10)$
and at $\kappa=0.81$ $R_{rg}=1.12(10)$. $R_{rg}$ seems to be greater
than one, but only marginally so, as the case $R_{rg}<1$ cannot be
excluded within error bars. On the other hand a look at
the second renormalization gives at $\kappa=0.79$, $R_{rg}=0.8$, now
the ratio indeed becoming smaller than 1.
In the crumpled phase we have no conclusive
evidence on the flow of the $\beta$ function.

\section{Conclusions}

The crumpling transition  separating
rigid and crumpled surfaces has been
object of intense scrutiny over the last decade.
In this paper we have presented a fresh approach
to the problem of determining the critical exponents associated
to that critical point. We have analyzed the transition using
Monte Carlo renormalization group techniques combining a
Fourier accelerated Langevin algorithm with a thinning of
degrees of freedom directly in momentum space. Where comparison
is possible  we have found good agreement with the
most recent simulations and found with great accuracy the value of
the fractal dimension at the critical point.
We have also given clear
evidence of the existence of a fixed point towards  which
all trajectories in the vicinity of the critical surface are driven
as well as the existence of a
unique relevant direction. We can confidently claim that our results
are  well established and have passed all the  necessary validity tests
to be trusted.

More computer resources should allow
a thorough investigation of a system of size
$128 \times 128$ where it should be possible to
perform three successive blockings. In the light of the present
experience as well as the one gained from  $\lambda\phi^4$ we believe
that such a simulation would  deliver the exponent $\nu$ with a 1\%
precision
(our level of precision here on $\nu$ is at the 10\% level). The precision
on $\eta$ could easily reach the per mille level (our current precision
on $\eta$ is 2\%).
Such a work would give very clear evidence about the nature of the
transition and with little additional effort
it would be possible to follow the renormalization group trajectories.
It would certainly
shed some light on some technical difficulties we encountered such as
the problems associated to the operators
\bref{badguys}. The interesting thing is that given the autocorrelation
times obtained using FALA such system is by no
means out of reach of existing computer facilities.
However, to remove some systematic uncertainties that we have
encountered the use of a hybrid algorithm in combination with blocking
in momentum space would be preferable.

Since the nature of the transition in three dimensions it is by now
well understood, an important question that remains open
is the nature of the transition in higher dimensions. In the first
studies of such models \cite{oldsim},\cite{am1} it was clearly
established that the transition gets weaker as the dimension
is increased, but nothing could be said about the order of the
transition for $D>4$ or the existence of an
upper critical dimension above which there is no transition at all.
To settle down these issues more work has to be done and certainly
MCRG will be a decisive instrument.

Another point of great interest would be to
implement such renormalization group
calculations to the case where we switch on gravity\cite{DTR1}. The
latest numerical simulations with such systems\cite{DTR2} suggest that
the transition becomes then of third order. Numerical simulation are
extremely time consuming for such models, and, up to know, there are
no reliable estimates for the critical exponents. It seems plausible
that a combination of $2d$ gravity renormalization group methods, like
the ones using self-similarity of the triangulations\cite{burda} and
the one described in this paper could shed some
 light on this very important
issue. Alternatively, Regge calculus with a non-local action
taking proper care of the conformal anomaly can also be
attacked by the same techniques. Work on this latter
approach is now under way.
\bigskip

{\bf Acknowledgements}

We would like to thank J. Wheater for discussions, correspondence
and for sharing some of his results with us. Discussions with
D.Johnston are also gratefully
acknowledged.
We would like to thank Ll. Garrido for expert advice on computing issues,
and CESCA and CEPBA for making their machines available to us.
A.T.
acknowledges a grant from the Generalitat de Catalunya. This
work has been supported in part by UE contract CHRX-CT93-0343 and
CICYT grant AEN93-0695.

\bigskip\newpage

\appendix

\section{Appendix}

In the continuum we classify operators according to their canonical
dimensions $d_c$. Recalling that $x$ is a dimensionless field in two
dimensions, a possible exhaustive list would be

$d_c=2$: $\nabla x \nabla x $.

$d_c=0$:
$Tr{K^2} $ and $R $.

$d_c= -2$:
$Tr{K^4} $, $Tr{K^3}Tr{K} $,
$(Tr{K^2})^2 $, $ Tr{\nabla K\nabla K} $,
$ (Tr{\nabla K})^2 $, $ R\ TrK^2 $ and $R^2$.
In the above expressions $K$ is the second fundamental form of
the surface and $R$ the gaussian curvature.
We did not consider operators odd in $x$ such as $tr K$ as they
are not present in the bare action and
will never be generated in the course
of a renormalization group transformation. We do not include
in our analysis any operator containing $R$
as we think that the transition is driven by terms
feeling the ambient space only.

\begin{figure}
\centerline{\psfig{figure=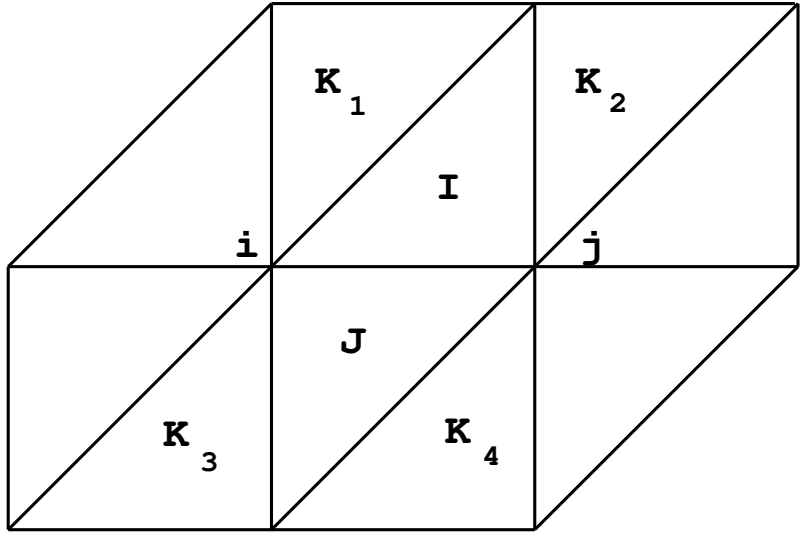,height=7cm}}
\caption{Conventions used in the discretizations of the operators.}
\label{fig:6}
\end{figure}

In order to discretize the above operators it is convenient
to introduce the following conventions. Given a site $i$,
$j(i)$ runs over the neighbouring sites. The same convention
applies to triangles, $J(I)$ runs over the triangles
adjacent to triangle $I$. Given a site $i$ and a nearest
neighbour $j(i)$ which selects a direction in our triangular
lattice with sixfold symmetry, we name the neighbouring triangles
in the manner depicted in fig. 6

\be
Tr K=\sum_{j(i)}\left(n_I - n_{K_1} \right)\cdot
      \left(x_j-x_i \right)
\ee
\be
Tr K^2=\frac{1}{2}\sum_{J(I)}
{\left( n_I - n_J \right)}^2=\sum_{J(I)}(1- n_I\cdot n_J)
\ee
\be
Tr K^3=\sum_{j(i)}\left(n_I^{\mu} n_{K_1}^{\nu}\right) \left(
x^{\nu}_j-x^{\nu}_{i} \right) \left(n_{K_2}^{\mu}+ n_{K_3}^{\mu}
\right)
\ee
\be
Tr K^4= \sum_{J(I)}(1-n_I\cdot n_J)^2
\ee
\be
(Tr \nabla K)^2=
\sum_{j(i)} {\left(\frac{1}{3}\left( n_{K_3}+ n_{K_4} \right)\cdot
\left (x_i- x_j \right) \right)}^2
\ee
\be
Tr \nabla K \nabla K = \Delta n_I \cdot \Delta n_I
\ee
where
the laplacian acting on $x$'s and $n$'s are defined by
\be
\Delta x_i=\sum_{j(i)}{\left(x_i-x_j
\right)}
\ee
\be
 \Delta n_I =  \left( n_I-
 \frac{1}{3}\sum_{J(I)} n_{J}
 \right)
\ee
These are the building blocks of our discretization.
It is important to ensure that symmetries of the bare action like
parity or discrete rotational symmetry
are preserved by the terms that we considered. To ensure the latter
we have symmetrized our operators when necessary.
The naming conventions of the different operators as appear in the
body of the paper are displayed in Table 14.

\begin{table}
\begin{center}
\begin{tabular}{cc}
$O_1$   &$ Tr K^2$                                \\
$O_2$   &$ Tr(K^2)^2$                             \\
$O_3$   &$ (TrK^2)^2$                             \\
$O_4$   &$ Tr(\nabla_a K \nabla_a K)$              \\
$O_5$   &$ Tr K Tr K^3 $                          \\
$O_6$   &$ (Tr \nabla K)^2 $                      \\
$O_7$   &$ {\left( (\nabla_a x)^2\right)}^2 $\\
$O_8$   &$ \Delta x \Delta  x $          \\
\end{tabular}
\end{center}
\caption{List of operators.}
\end{table}

While the discretization of some operators (in particular those involving
only normals) is rather natural, this is not so in other cases. To
get a feeling of the errors involved in the discretization procedure
we have also tried the following ones
\be
Tr K =\sum_{j(i)}{\left(n_{K_1} -n_{K_2}\right)\cdot \left(x_i
-x_j \right)}
\ee
\be
Tr K^3=\left(\sum_j \left(x_i^{\mu}-x_j^{\mu}
\right) \left(n_{K_1}^{\nu} -n_{K_2}^{\nu} \right) \right) \left(
\sum_{j} \left(n_{K_1}^{\mu}-n_{K_2}^{\mu}\right) \left(
n_{K_1}^{\nu}-n_{K_2}^\nu \right) \right)
\ee
We have changed accordingly $O_5$ and $O_6$.

The results show only very slight modifications. For instance after
including the new discretizations we get a similar pattern of
eigenvalues with no significative improvement except perhaps that some
of the imaginary parts become actually smaller with the new discretization
and the dispersion of eigenvalues is slightly reduced. For instance
in a $32\times 32\to 16\times 16\to 8\times 8$ blocking at $\kappa=0.82$
we got with $6 \times 10^5$ configurations the
sequence of eigenvalues: 2.08, 2.07, 2.21, 2.19, 2.18, 2.19,
1.96, 1.92, whose average is 2.10.
(to be compared with the average 2.17 in Table 10).

\vfill
\eject

\end{document}